\documentclass[aps,prl,english,amsmath,amsfonts,amssymb,superscriptaddress,twocolumn,showpacs,citeautoscript,reprint,longbibliography]{revtex4-1}
\usepackage[T1]{fontenc}
\usepackage[latin9]{inputenc}
\usepackage{amsmath}
\usepackage{amssymb}
\usepackage{wasysym}
\usepackage{adjustbox}
\usepackage{esint}
\usepackage{graphicx}
\usepackage{times}
\usepackage{nicefrac}
\usepackage{appendix}
\usepackage{multirow}
\usepackage{textcase}
\usepackage{subfigure}
\usepackage{empheq}
\usepackage{color}
\usepackage{leftidx}
\usepackage{hhline}
\usepackage{makecell}
\usepackage{stackrel}
\usepackage{lineno}
\makeatletter
\@ifundefined{textcolor}{}
{%
 \definecolor{BLACK}{gray}{0}
 \definecolor{WHITE}{gray}{1}
 \definecolor{RED}{rgb}{1,0,0}
 \definecolor{GREEN}{rgb}{0,1,0}
 \definecolor{BLUE}{rgb}{0,0,1}
 \definecolor{CYAN}{cmyk}{1,0,0,0}
 \definecolor{MAGENTA}{cmyk}{0,1,0,0}
 \definecolor{YELLOW}{cmyk}{0,0,1,0}
}

\renewcommand{\vec}[1]{\mathbf{#1}}

\renewcommand{\Im}{\operatorname{Im}}

\renewcommand{\b}{\beta}

\newcommand{\add}[1]{\if\a\b{{\color{red} #1}}\else{#1}\fi}

\newcommand{\bracket}[1]{\langle #1 \rangle}
\newcommand{\ket}[1]{| #1 \rangle}
\newcommand{\bra}[1]{\langle #1 |}
\newcommand{\im}{\operatorname{i}}

\renewcommand{\eqref}[1]{(\ref{eq:#1})}

\newcommand{\figref}[1]{Fig.~\ref{fig:#1}}
\newcommand{\Figref}[1]{Figure~\ref{fig:#1}}

\newcommand{\TT}{\mathbb{T}}

\newcommand{\VV}{\mathbb{V}}

\newcommand{\II}{\mathbb{I}}

\newcommand{\Gvac}{\mathbb{G}^{\mathrm{vac}}}
\newcommand{\Gsca}{\mathbb{G}^{\mathrm{sca}}}

\makeatother
\usepackage{babel}

\begin{document}

\title{Fundamental limits to attractive and repulsive Casimir--Polder forces}

\author{Prashanth S. Venkataram}
\author{Sean Molesky}
\author{Pengning Chao}
\author{Alejandro W. Rodriguez}
\affiliation{Department of Electrical Engineering, Princeton
  University, Princeton, New Jersey 08544, USA}

\date{\today}

\begin{abstract}
  We derive upper and lower bounds on the Casimir--Polder force
  between an anisotropic dipolar body and a macroscopic body separated
  by vacuum via algebraic properties of Maxwell's equations. These
  bounds require only a coarse characterization of the system---the
  material composition of the macroscopic object, the polarizability
  of the dipole, and any convenient partition between the two
  objects---to encompass all structuring possibilities. We find that
  the attractive Casimir--Polder force between a polarizable dipole
  and a uniform planar semi-infinite bulk medium always comes within
  10\% of the lower bound, implying that nanostructuring is of limited
  use for increasing attraction. In contrast, the possibility of
  repulsion is observed even for isotropic dipoles, and is routinely
  found to be several orders of magnitude larger than any known
  design, including recently predicted geometries involving conductors
  with sharp edges. Our results have ramifications for the design of
  surfaces to trap, suspend, or adsorb ultracold gases.
\end{abstract}

\maketitle 

Casimir--Polder (CP) forces between polarizable dipolar bodies and
macroscopic objects arise from zero-point fluctuations of the
electromagnetic field~\cite{Buhmann2012I, Buhmann2012II,
  Intravaia2011, DeKieviet2011, AgarwalPRA1975, VenkataramPRL2017,
  BarcellonaPRA2016, WoodsRMP2015, TkatchenkoPRL2012,
  ChattopadhyayaCM2017, HermannCR2017}. These forces, along with more
general Casimir forces, have been experimentally measured in systems
including planar substrates, gratings, Rydberg atoms, molecules, and
Bose--Einstein condensates~\cite{Buhmann2012I, Buhmann2012II,
  Intravaia2011, DeKieviet2011, WagnerNATURE2014, BenderPRX2014,
  KlimchitskayaRMP2009}. Owing to their strong and complicated
dependence on geometry, prior works have sought means of modifying the
magnitude and sign of these forces (beyond the typical attractive and
monotonically decaying power laws~\cite{ZarembaPRB1976,
  AgarwalPRA1975, Buhmann2012I, Buhmann2012II, Intravaia2011}) via
nanostructuring. In particular, outside of systems satisfying the
Dzyaloshinskii--Lifshitz--Pitaevskii permittivity criterion for
repulsion~\cite{Dzyaloshinskii1961, MundayNATURE2009, ZhaoSCIENCE2019,
  MiltonJPA2012, PirozhenkoPRA2009} (requiring an intervening medium,
e.g. fluids), repulsive Casimir and CP forces have been predicted for
bodies separated in vacuum mainly for anisotropic dipoles at small
separations~\cite{ShajeshPRA2012, MiltonPRD2012}, planar magnetic
media~\cite{PirozhenkoPRA2009, RosaPRD2010, PirozhenkoJPA2008,
  LambrechtPRA2008}, metallic rectangular
gratings~\cite{BuhmannIJMPA2016}, metallic or dielectric plates with
circular holes~\cite{LevinPRL2010, EberleinPRA2011, ShajeshPRA2012},
and other metallic surfaces with sharp edges~\cite{MaghrebiPRD2011,
  MiltonPRA2011}. Likewise, beyond the general no-go theorem for
repulsion in mirror-symmetric systems in vacuum~\cite{KennethPRL2006}
or recent generalizations of Earnshaw's theorem setting constraints on
stable equilibria~\cite{RahiPRL2010}, quantitative limits on
attractive or repulsive Casimir forces have mainly been restricted to
uniform planar dielectric and magnetic media~\cite{HenkelEPL2005,
  LambrechtPLA1997}. Understanding bounds on CP forces is crucial for
designing surfaces to trap, adsorb, or suspend atoms, molecules, and
quantum emitters~\cite{Buhmann2012I, Buhmann2012II, Intravaia2011,
  KlimchitskayaRMP2009}.

\begin{figure}[t!]
\centering
\includegraphics[trim=40 60 40 0,clip,width=0.95\columnwidth]{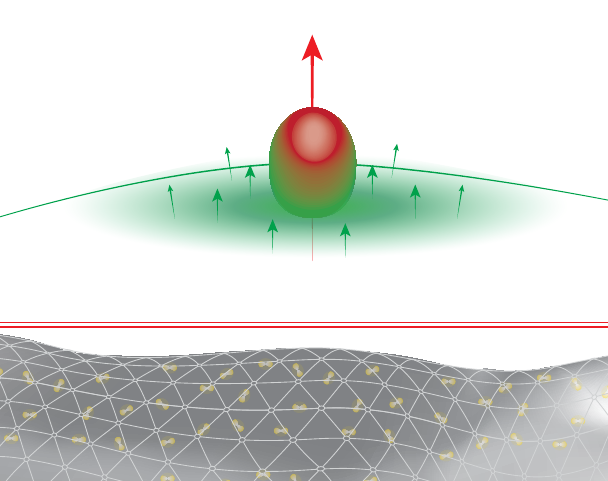}
\caption{\textbf{Schematic of investigation}. We derive
  shape-independent upper and lower bounds on the Casimir--Polder
  force between a polarizable dipolar body of parallel (perpendicular)
  polarizability $\alpha_\parallel$ ($\alpha_\perp$) above any
  nanostructured medium of susceptibility $\chi$ within a given
  domain.}
\label{fig:schem}
\end{figure}

In this paper, we present upper and lower bounds on CP forces for a
dipolar body separated by vacuum from a macroscopic body of uniform
susceptibility that depend only on the dipole polarizability, the
susceptibility of the macroscopic body, and the choice of a partition
separating the two objects, shown schematically in~\figref{schem};
positive bounds correspond to repulsion, while negative bounds
correspond to attraction. Surprisingly, these simple properties
capture all of the physical properties needed for the
bounds. Regardless of anisotropy, the archetypal CP force between a
dipole and a semi-infinite planar bulk is consistently within 10\% of
the lower bound; thus, for attraction, the bound is relatively tight
and nanostructuring can offer only modest improvements. Conversely,
sharp contrasts between the bounds and known designs are observed for
repulsive forces. Regardless of the polarizability of the dipolar
body, repulsion is never completely ruled out, and in most cases the
bound is found to be several orders of magnitude larger than what has
been observed in any known design, including recently proposed special
geometries involving highly anisotropic dipolar bodies and metals with
sharp edges~\cite{LevinPRL2010, EberleinPRA2011, ShajeshPRA2012} which
prove challenging to probe experimentally. This finding suggests that
nontrivial nanostructuring may yet lead to practically feasible
designs with strong repulsive CP forces~\footnote{The relatively weak
  forces and large dipolar anisotropies needed to observe repulsion in
  the geometry of a needle above a plate with a hole poses challenges
  for experimental observation~\cite{LevinPRL2010}.}. Since the
magnitudes of both the attractive and repulsive bounds grow increasing
domain size, in what follows we focus our attention on structures
contained within a planar semi-infinite half-space.

As notation, a vector field $\vec{v}(\vec{x})$ will be denoted as
$\ket{\vec{v}}$. At $\omega = \im\xi$, all relevant polarization and
field quantities can be defined to be real-valued in position space
without loss of generality, so we define the \emph{unconjugated} inner
product $\bracket{\vec{u}, \vec{v}} = \int~\mathrm{d}^{3} x~\vec{u}
(\vec{x}) \cdot \vec{v}(\vec{x})$. An operator $\mathbb{A}(\vec{x},
\vec{x}')$ will be denoted as $\mathbb{A}$, with $\int~\mathrm{d}^{3}
x'~\mathbb{A}(\vec{x}, \vec{x}') \cdot \vec{v}(\vec{x}')$ denoted as
$\mathbb{A}\ket{\vec{v}}$.
Unless stated otherwise, all quantities are taken to implicitly depend
on $\im\xi$.

\emph{Main result.}---Consider the CP force on a dipole of
susceptibility $\VV_{\mathrm{dipole}} = \sum_{\beta} \alpha_{\beta}
\ket{\vec{u}^{(\beta)}} \bra{\vec{u}^{(\beta)}}$, where the
polarizabilities $\alpha_{\beta}(\im\xi)$ are positive and the basis
functions $\vec{u}^{(\beta)}(\vec{x}) = \vec{n}_{\beta} \delta^{3}
(\vec{x} - \vec{R})$ are given in terms of the dipole location
$\vec{R}$ and the principal axes $\vec{n}_{\beta}$. The upper and
lower bounds (repulsion and attraction) on the CP force along
Cartesian axis $k$, respectively $F^{+}_{k}$ and $F^{-}_{k}$, are
shown to depend on $\VV_{\mathrm{dipole}}$, the macroscopic
susceptibility $\chi(\im\xi)$ (assumed to be homogeneous, local, and
isotropic), and the choice of domain enclosing the macroscopic body,
all of which are completely general, independent of any particular
material dispersion model or body shapes. As argued in the following
derivation, the lower bounds can never increase, and the upper bounds
never decrease, when the chosen domain is enlarged, so that as a whole
the bounds are domain monotonic. Based on this fact, the bounds on the
CP force can be written as
\begin{multline} \label{eq:CPforcebounds}
  F^{\pm}_{k} = \frac{\hbar}{2\pi} \int_{0}^{\infty} \sum_{\beta}
  \alpha_{\beta} \Bigg[\left\langle\frac{\partial
      \vec{u}^{(\beta)}}{\partial R_{k}}, \Gsca
    \vec{u}^{(\beta)}\right\rangle \pm
    \\ \left(\left\langle\vec{u}^{(\beta)}, \Gsca
    \vec{u}^{(\beta)}\right\rangle \left\langle\frac{\partial
      \vec{u}^{(\beta)}}{\partial R_{k}}, \Gsca \frac{\partial
      \vec{u}^{(\beta)}}{\partial R_{k}}\right\rangle\right)^{1/2}
    \Bigg]~\mathrm{d}\xi,
\end{multline}
where again all quantities are evaluated at $\omega = \im\xi$. Here,
$\Gsca$ is the scattering Green's function of the equivalent object
created by filling the domain in question \emph{uniformly} with a
material of susceptibility $\chi(\im\xi)$, and \emph{not} the
scattering Green's function of any object possibly contained within
the domain. (Crucially, \eqref{CPforcebounds} is not the expression of
the CP force for any particular geometry.) These bounds need not have
a definite sign either: for combinations of dipole polarizability,
position and domain choice where the upper bound is positive and the
lower bound is negative, there are potentially structures producing
either attractive or repulsive CP forces.

\emph{Technical Derivation.}---The CP force $F_{k} =
\frac{\partial}{\partial R_{k}} \int \bracket{\vec{E}(t, \vec{x})
  \cdot \vec{P}(t, \vec{x})}~\mathrm{d}^{3} x$~\cite{AgarwalPRA1975,
  RosaPRA2011, VenkataramPRL2017, HermannCR2017} by a macroscopic body
of susceptibility $\VV$ on a dipole of susceptibility
$\VV_{\mathrm{dipole}}$ as above can be derived briefly as follows,
where $\bracket{\ldots}$ refers to a thermodynamic average. We take
the Fourier transform to real $\omega$ and solve the integral form of
Maxwell's equations $\ket{\vec{E}} = \ket{\vec{E}^{(0)}} +
\Gvac\ket{\vec{P}}$ simultaneously with $\ket{\vec{P}} =
\ket{\vec{P}^{(0)}} + (\VV + \VV_{\mathrm{dipole}})\ket{\vec{E}}$ for
$\ket{\vec{E}}$ and $\ket{\vec{P}}$ for real $\omega$, where $\Gvac$
solves $(\nabla \times (\nabla \times) - (\omega/c)^{2} \II)\Gvac =
-(\omega/c)^{2} \II$. We define $\TT = (\II - \VV\Gvac)^{-1} \VV$ for
the macroscopic body, which has the same support as $\VV$ and
satisfies $\TT = \TT(\VV^{-1} - \Gvac)\TT$, where $\Gvac$ is
implicitly projected onto a domain which contains the support of
$\VV$. Finally, we use the zero-temperature fluctuation--dissipation
relations $\bracket{\ket{\vec{E}^{(0)}(\omega)}
  \bra{\vec{E}^{(0)}(\omega')}} = \hbar\Im(\Gvac(\omega)) \times
2\pi\delta(\omega - \omega')$ and
$\bracket{\ket{\vec{P}^{(0)}(\omega)} \bra{\vec{P}^{(0)}(\omega')}} =
\hbar\Im(\VV(\omega) + \VV_{\mathrm{dipole}}(\omega)) \times
2\pi\delta(\omega - \omega')$, perform a Wick rotation to $\omega =
\im\xi$ as the integrand is analytic in the upper-half plane of
$\omega$, and then expand to lowest order in scattering between the
dipole and macroscopic body to yield
\begin{equation}
  F_{k} = \frac{\hbar}{2\pi} \int_{0}^{\infty} \sum_{\beta}
  \alpha_{\beta} \frac{\partial}{\partial R_{k}} \left\langle
  \vec{u}^{(\beta)}, \Gvac\TT\Gvac \vec{u}^{(\beta)}
  \right\rangle~\mathrm{d}\xi
\end{equation}
as the CP force. Henceforth, we assume that $\VV$ represents a scalar
(homogeneous, local, isotropic) susceptibility $\chi$. Our goal then
is to find bounds such that $F_{k} \in [F^{-}_{k}, F^{+}_{k}]$. We
note that at $\omega = \im\xi$, $\VV$, $\Gvac$, and $\TT$ in general
are real-symmetric operators in position space, with $\VV$ and $\TT$
being positive-definite while $\Gvac$ is negative-definite (and this
applies to its diagonal projected blocks too).

We first consider the problem of optimizing $\frac{\partial}{\partial
  R_{k}} \bracket{\vec{E}^{\mathrm{inc}}, \TT\vec{E}^{\mathrm{inc}}} =
2\left\langle \frac{\partial \vec{E}^{\mathrm{inc}}}{\partial R_{k}},
\TT\vec{E}^{\mathrm{inc}} \right\rangle$ for an arbitrary incident
field $\ket{\vec{E}^{\mathrm{inc}}}$. In particular, we define the
action of $\TT$ to be a new vector $\ket{\vec{P}} =
\TT\ket{\vec{E}^{\mathrm{inc}}}$, and optimize the quantity
$2\left\langle \frac{\partial \vec{E}^{\mathrm{inc}}}{\partial R_{k}},
\vec{P} \right\rangle$ with respect to $\ket{\vec{P}}$, assuming that
the response $\ket{\vec{P}}$ can be chosen arbitrarily given its
support. However, we also take care to impose the equality constraint
$\TT = \TT(\VV^{-1} - \Gvac)\TT$ to ensure physical consistency:
evaluating this with respect to $\ket{\vec{E}^{\mathrm{inc}}}$ gives
$\bracket{\vec{E}^{\mathrm{inc}}, \vec{P}} = \chi^{-1}
\bracket{\vec{P}, \vec{P}} - \bracket{\vec{P}, \Gvac\vec{P}}$, and
this quantity is positive as $\TT$ is positive-definite. For
convenience, we define the eigenvalue decomposition of the projection
of $\Gvac$ into the given domain as $\Gvac = -\sum_{\mu} \rho_{\mu}
\ket{\vec{N}^{(\mu)}} \bra{\vec{N}^{(\mu)}}$, where $\rho_{\mu} > 0$
and $\bracket{\vec{N}^{(\mu)}, \vec{N}^{(\nu)}} = \delta_{\mu\nu}$,
and define the basis expansions $v_{\mu} = \bracket{\vec{N}^{(\mu)},
  \vec{E}^{\mathrm{inc}}} = \bracket{\vec{E}^{\mathrm{inc}},
  \vec{N}^{(\mu)}}$ and $t_{\mu} = \bracket{\vec{N}^{(\mu)}, \vec{P}}
= \bracket{\vec{P}, \vec{N}^{(\mu)}}$. As the domain choice is
independent of $\vec{R}$, then $2\left\langle \frac{\partial
  \vec{E}^{\mathrm{inc}}}{\partial R_{k}}, \vec{P} \right\rangle =
2\sum_{\mu} \frac{\partial v_{\mu}}{\partial R_{k}} t_{\mu}$. This
leads to the constrained optimization of the objective
\begin{equation}
  L = \sum_{\mu} \left[2\frac{\partial v_{\mu}}{\partial R_{k}} t_{\mu} -
  \lambda (t_{\mu} v_{\mu} - (\chi^{-1} + \rho_{\mu})t_{\mu}^{2})\right]
\end{equation}
where $\lambda$ is a Lagrange multiplier. As we have chosen the domain
into which we project $\Gvac$ to contain the support of
$\ket{\vec{P}}$---encoded in the expansion coefficients
$\{t_{\mu}\}$---enlarging the domain into which we project $\Gvac$
cannot affect the equality constraint. Similarly, the magnitude of the
objective cannot decrease with increasing domain because
$\ket{\vec{P}}$ can access the smaller domain. If no better
performance is possible, $\ket{\vec{P}}$ can always be taken to be the
previous solution. Thus, our bound is domain monotonic, and so any
domain with projection operator $\mathbb{I}_\mathrm{d}$ that fully
encloses all possible object designs of interest can be used to
generate bounds.

Carrying out the optimization yields the equations $2\frac{\partial
  v_{\mu}}{\partial R_{k}} - \lambda(v_{\mu} - 2(\chi^{-1} +
\rho_{\mu}) t_{\mu}) = 0$ and $\sum_{\mu} (t_{\mu} v_{\mu} -
(\chi^{-1} + \rho_{\mu})t_{\mu}^{2}) = 0$. The first equation gives
$t_{\mu} = \frac{1}{\chi^{-1} + \rho_{\mu}} \left(\frac{v_{\mu}}{2} -
\frac{1}{\lambda} \frac{\partial v_{\mu}}{\partial R_{k}}\right)$, and
plugging this into the second equation gives
\[
\lambda \in \pm
2\sqrt{\frac{\left\langle \frac{\partial
      \vec{E}^{\mathrm{inc}}}{\partial R_{k}}, (\chi^{-1}
    \II_{\mathrm{d}} - \Gvac)^{-1} \frac{\partial
      \vec{E}^{\mathrm{inc}}}{\partial R_{k}}
    \right\rangle}{\left\langle\vec{E}^{\mathrm{inc}}, (\chi^{-1}
    \II_{\mathrm{d}} - \Gvac)^{-1}
    \vec{E}^{\mathrm{inc}}\right\rangle}}.
\] 
The constrained objective has $\frac{\delta^{2} L}{\delta t_{\mu}
  \delta t_{\nu}} = 2\lambda (\chi^{-1} + \rho_{\mu})
\delta_{\mu\nu}$, so the negative value of $\lambda$ gives the maximum
while the positive value gives the minimum. (Another special
stationary point corresponding to $\lambda = 0$, which is a saddle
point, can be found if $\ket{\frac{\partial
    \vec{E}^{\mathrm{inc}}}{\partial R_{k}}} = 0$; this cannot arise
for the incident field radiated by a dipole into a domain, so we do
not consider it further.) Hence, $L \in [L^{-}, L^{+}]$, with
\begin{widetext}
\begin{equation}
  L^{\pm} = \left\langle \frac{\partial
    \vec{E}^{\mathrm{inc}}}{\partial R_{k}}, (\chi^{-1}
  \II_{\mathrm{d}} - \Gvac)^{-1} \vec{E}^{\mathrm{inc}}
  \right\rangle \pm \sqrt{\left\langle \frac{\partial
      \vec{E}^{\mathrm{inc}}}{\partial R_{k}}, (\chi^{-1}
    \II_{\mathrm{d}} - \Gvac)^{-1} \frac{\partial
      \vec{E}^{\mathrm{inc}}}{\partial R_{k}}
    \right\rangle\left\langle \vec{E}^{\mathrm{inc}}, (\chi^{-1}
    \II_{\mathrm{d}} - \Gvac)^{-1} \vec{E}^{\mathrm{inc}}
    \right\rangle}.
\end{equation}
\end{widetext}
For our problem of interest we set $\ket{\vec{E}^{\mathrm{inc}}} =
\Gvac\ket{\vec{u}^{(\beta)}}$ and identify $\Gsca = \Gvac (\chi^{-1}
\II_{\mathrm{d}} - \Gvac)^{-1} \Gvac$ as the scattering Green's
function of the equivalent object formed by filling the entire domain
of interest with the susceptibility $\chi$. As each $\alpha_{\beta}
(\im\xi) > 0$, the net upper bound cannot rise above the upper bound
applied to each channel $\beta$, just as the net lower bound cannot
fall below the per-channel lower bound.  This argument also applies to
integration over $\xi$ and so ~\eqref{CPforcebounds} follows.

\emph{Discussion.}---Domain monotonicity allows us to choose the
largest domain enclosing any desired design. As gratings, plates with
apertures, wedges, and knife-edge geometries have all been studied in
the context of CP repulsion, and since experiments typically consider
extended nanostructured media, we take the domain to be a planar
semi-infinite half-space; this choice ensures the existence of a
separating plane between the dipole and the macroscopic object, in
contrast to interleaved geometries~\cite{RodriguezPRA2008} allowing
effective repulsion through lateral forces. For this choice, $\Gsca$
of the equivalent object admits semianalytical
expressions~\cite{Novotny2006}. Additionally, the CP force and its
bounds are linear functionals of the polarizabilities ${\alpha_{\beta}
  (\im\xi)}$, and become simple linear functions if the
polarizabilities are assumed to be dispersionless. Thus, for
simplicity, we consistently choose the principal axes to align with
the Cartesian axes, and consider $\alpha_{x}(0) = \alpha_{y}(0) =
\alpha_{\parallel}$ and $\alpha_{z}(0) = \alpha_{\perp}$. The dipole
location is taken to be $\vec{R} = d\vec{e}_{z}$, where $d$ is the
minimum separation of the dipole from the design domain, and the force
direction of interest to lie along $\vec{e}_{z}$. This choice of
polarizabilities allows us to decompose the force into parallel and
perpendicular components, $F_{z}^{\pm} = g^{\pm}_{\parallel} +
g^{\pm}_{\perp} \alpha_{\perp}/\alpha_{\parallel}$ for appropriate
functions $g^{\pm}_{\parallel}$ and $g^{\pm}_{\perp}$ which are
linearly proportional to $\alpha_{\parallel}$, so
$F_{z}^{\pm}/\alpha_{\parallel}$ is an affine linear function of the
polarizability ratio (similar to an aspect ratio)
$\alpha_{\perp}/\alpha_{\parallel}$. These assumptions make evaluation
and analysis of the CP force bounds particularly convenient.

\begin{figure}[t!]
\centering
\includegraphics[width=0.95\columnwidth]{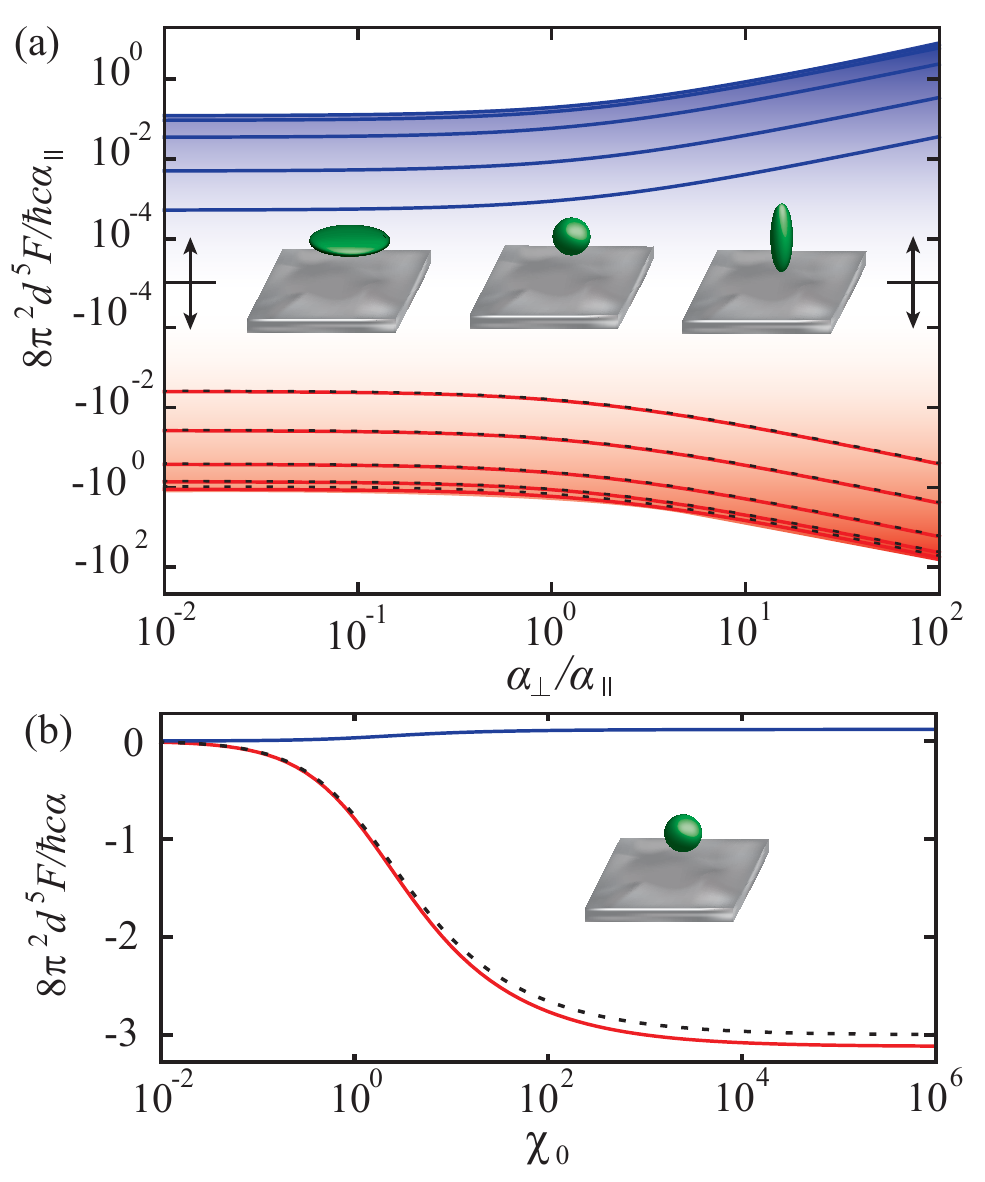}
\caption{\textbf{Material and anisotropy dependence of bounds on
    Casimir--Polder forces}. (a) Upper and lower bounds to the CP
  force (blue and red lines) on a nondispersive anisotropic dipole of
  parallel and perpendicular polarizabilities $\alpha_{\parallel}$ and
  $\alpha_{\perp}$, above a planar semi-infinite half-space domain,
  along with the actual force above a planar semi-infinite bulk
  (dashed lines), normalized to $\hbar c\alpha_{\parallel}/8\pi^{2}
  d^{5}$ for separation $d$, as a function of
  $\alpha_{\perp}/\alpha_{\parallel}$. The macroscopic susceptibility
  $\chi_{0}$ is nondispersive and increases logarithmically from
  $10^{-2}$ to $10^{6}$ (lighter to darker shades). (b) Same as (a)
  but plotted against $\chi_{0}$ for the isotropic case
  $\alpha_{\perp} = \alpha_{\parallel}$.}
\label{fig:chi0fixed}
\end{figure}

We begin by considering a macroscopic body of dispersionless
susceptibility $\chi(\im\xi) = \chi_{0}$. This leads to the simple
result that the bounds $F_{z}^{\pm}$ for this domain, as well as the
CP force for a nondispersive dipole above a planar semi-infinite bulk
of susceptibility $\chi_{0}$, both scale as $d^{-5}$. Therefore, we
need only consider the dependence of these bounds on $\chi_{0}$ as
well as the polarizability ratio
$\alpha_{\perp}/\alpha_{\parallel}$. \Figref{chi0fixed} shows the
bounds, as well as the actual attractive CP force above a planar
semi-infinite bulk of susceptibility $\chi_{0}$, as a function of
$\alpha_{\perp}/\alpha_{\parallel}$ for multiple $\chi_{0}$ (a), and
as a function of $\chi_{0}$ for the isotropic case $\alpha_{\perp} =
\alpha_{\parallel}$ (b). As expected, for any nonzero $\chi_{0}$ and
$\alpha_{\parallel}$, the bounds and planar force (normalized by the
dependence on $d$ and $\alpha_{\parallel}$) attain a nonzero value for
$\alpha_{\perp} = 0$, and increase linearly with
$\alpha_{\perp}/\alpha_{\parallel}$; moreover, the bounds increase
monotonically with $\chi_{0}$, saturating at finite values in the
perfect electrically conducting (PEC) limit $\chi_{0} \to
\infty$. Stunningly, the actual force is consistently within 10\% of
the lower bound for all $\chi_{0}$ and
$\alpha_{\perp}/\alpha_{\parallel}$, indicating that nanostructuring
can only weakly enhance attractive CP forces in extended
geometries. In~\eqref{CPforcebounds}, the first term in the summand is
half of the actual force above a planar semi-infinite bulk, so the
second term is crucial to making the bounds valid and tight for this
domain choice. Conversely, at every $\chi_{0}$ and
$\alpha_{\perp}/\alpha_{\parallel}$, the upper bound is positive,
suggesting that at any $d$ and for any polarizability ratio and
$\chi_{0}$, there are in fact potential macroscopic geometries that
can meaningfully repel dipoles. The tightness of the lower bounds
indicates that these limits capture essential physics. Hence, it is
fairly plausible that tailored macroscopic geometries approaching the
upper bound do exist. It is also worth mentioning that in the few
geometries where repulsion is predicted for strongly anisotropic
dipoles, it is prohibited for isotropic dipoles, but our upper bounds
do not rule out the existence of other repulsive geometries even for
isotropic dipoles. Finally, we point out that the magnitude of the
upper bounds are consistently more than an order of magnitude smaller
than the magnitude of the lower bounds.

\begin{figure}[t!]
\centering
\includegraphics[width=0.95\columnwidth]{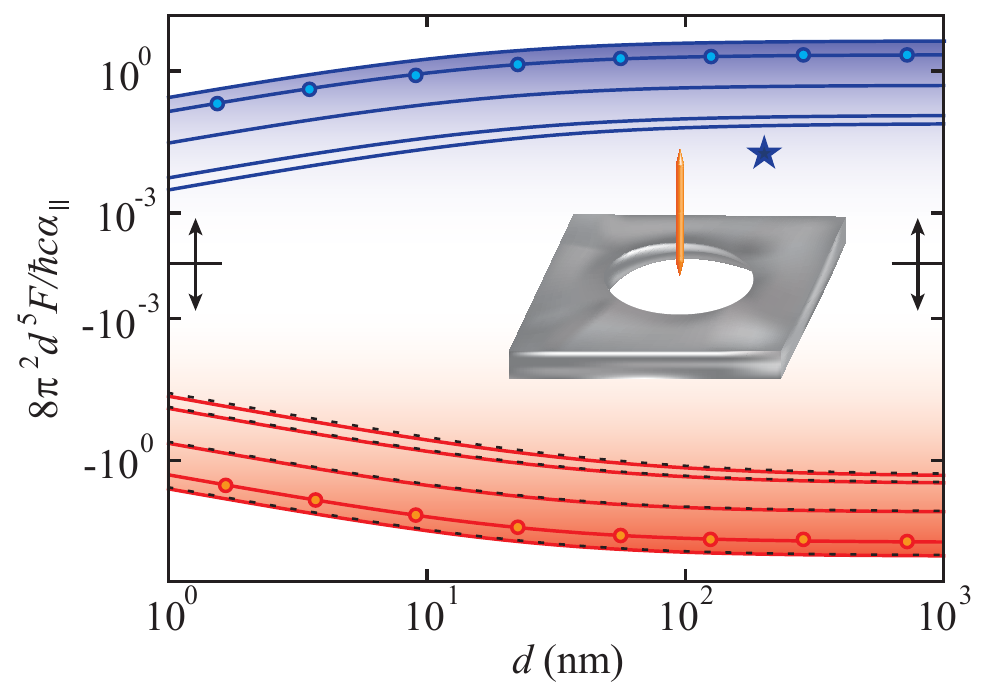}
\caption{\textbf{Distance dependence of bounds on Casimir--Polder
    forces for gold nanostructures}. Repulsive and attractive bounds
  on CP forces as in~\figref{chi0fixed} but with the macroscopic
  susceptibility $\chi$ corresponding to that of gold and
  $\alpha_{\perp}/\alpha_{\parallel}$ increasing logarithmically from
  $10^{-2}$ to $10^{2}$ (lighter to darker shades). Also shown is the
  CP force on a gold needle above a gold plate with a circular
  aperture from Ref.~\cite{LevinPRL2010} (dark blue star),
  corresponding to a static anisotropic polarizability ratio
  $\alpha_{\perp}/\alpha_{\parallel} \approx 51.1$ at $d =
  200~\mathrm{nm}$; bounds for $\alpha_{\perp}/\alpha_{\parallel} =
  50$ are marked in lines with circles.}
\label{fig:chiAu}
\end{figure}

Next, we relax the assumption that $\chi(\im\xi)$ is nondispersive,
and consider the particular case of a gold medium, for which
$\chi(\im\xi) = \omega_{\mathrm{p}}^{2} / (\xi^{2} + \gamma\xi)$ for
$\omega_{\mathrm{p}} = 1.37\times 10^{16}~\mathrm{rad/s}$ and $\gamma
= 5.32\times 10^{13}~\mathrm{rad/s}$. For simplicity, we continue to
neglect dispersion in $\alpha_{\parallel}$ and $\alpha_{\perp}$, so
the linear scaling of the bounds with $\alpha_{\parallel}$ and affine
linear scaling with $\alpha_{\perp}/\alpha_{\parallel}$ are
preserved. The introduction of dispersion means that the bounds no
longer scale uniformly as $d^{-5}$: as seen in~\figref{chiAu}, the
bounds transition from the nonretarded scaling of $d^{-4}$ toward
$d^{-5}$ as the separation increases. The linear increase in the
bounds with $\alpha_{\perp}/\alpha_{\parallel}$ is also clear. More
importantly, while more than an order of magnitude smaller than the
lower bounds, for a dispersive metal like gold the possibility of
repulsion is still not ruled out as the upper bounds remain positive
for all $d$ and $\alpha_{\perp}/\alpha_{\parallel}$. For attraction,
the actual forces produce from the planar geometry are again within
10\% of the corresponding lower bounds for all $d$ and
$\alpha_{\perp}/\alpha_{\parallel}$, demonstrating that these results
are not simply artifacts of a nondispersive $\chi$. We further compare
the bounds for gold to the actual repulsive force by a gold plate with
a circular aperture upon a gold nanorod at a center-to-center
separation of $d = 200~\mathrm{nm}$, approximating the nanorod as an
ellipsoid with the same major and minor axes ($320~\mathrm{nm}$ and
$20~\mathrm{nm}$, respectively) using the anisotropic
Clausius--Mossotti form of the polarizability~\cite{Bohren2007CH5}.
The dipolar approximation may not be valid given that the separation
is smaller than the major axis length, but we use this simply as a
heuristic to make qualitative comparisons to our bounds. Approximating
the nanorod as a PEC, the polarizability ratio is
$\alpha_{\perp}/\alpha_{\parallel} \approx 51.1$: for that ratio and
$d$, the actual force~\cite{LevinPRL2010} is more than 2 orders of
magnitude smaller than the upper bound, strongly suggesting that
macroscopic geometries optimized for CP repulsion may look quite
different from prior proposed geometries. Finally, we note that the
bounds show qualitatively similar behavior for polar dielectrics like
undoped silicon as for metals like gold, though those of silicon are
smaller than their counterparts for gold.

In summary, we have derived bounds for the CP force on a general
anisotropic dipolar body by a macroscopic body of susceptibility
$\chi$ enclosed within a prescribed domain, and have evaluated these
bounds specifically for a planar semi-infinite half-space domain. The
lower bounds are nearly achieved by the typical geometry of a dipole
above a uniform planar body, whereas existing predictions of repulsive
CP forces in geometries involving conductors with sharp edges fall
nearly two orders of magnitude below the limits on repulsion. We
expect that similar to other nanophotonic phenomena like local density
of states modifications and radiative heat transfer, optimal
structures for attraction or repulsion found through brute-force
techniques such as inverse design~\cite{JinPRB2019, JinOE18,
  MoleskyNATURE2018} will look very different from the high-symmetry
geometries proposed thus far, and that the tightness of the lower
bounds for known structures suggests that appropriately designed
structures may indeed approach the upper bounds and yield measurable
repulsive CP forces even for relatively isotropic dipoles like Rydberg
atoms, in contrast to existing designs~\cite{LevinPRL2010,
  EberleinPRA2011, ShajeshPRA2012, Buhmann2012I,
  Buhmann2012II}. Additionally, we point out that the Casimir energy
between a dipolar particle and a macroscopic object in vacuum is
always negative and goes to zero at asymptotically large separations,
precluding a macroscopic geometry that repels a dipole for every
location (as the force must be attractive sufficiently far away); this
suggests that the upper bounds could be further tightened. Though our
results focused on the force normal to the plane separating the
dipolar and extended bodies, \eqref{CPforcebounds} can be employed to
bound lateral forces, the subject of much recent
interest~\cite{ChenPRL2002, EmigPRA2003, KlimchitskayaRMP2009,
  KlimchitskayaJPCS2010, ManjavacasPRL2017}, as well as forces
involving compact objects. Finally, we point out that our bounds can
be easily generalized to finite temperature equilibrium CP forces by
replacing the frequency integration with a Matsubara
summation~\cite{Buhmann2012II, Intravaia2011}.

\emph{Acknowledgments.}---This work was supported by the National
Science Foundation under Grants No. DMR-1454836, DMR 1420541, DGE
1148900, the Cornell Center for Materials Research MRSEC (award
no. DMR1719875), and the Defense Advanced Research Projects Agency
(DARPA) under agreement HR00111820046. The views, opinions and/or
findings expressed herein are those of the authors and should not be
interpreted as representing the official views or policies of any
institution.

\nocite{apsrev41Control} \bibliographystyle{apsrev4-1}

\bibliography{caspolboundspaper}
\end{document}